\documentclass{aa}
\usepackage[varg]{txfonts}
\usepackage{booktabs}
\usepackage{graphicx}
\usepackage{xcolor}
\usepackage[colorlinks=true,linkcolor=blue,citecolor=blue,filecolor=blue,urlcolor=blue]{hyperref}

\newcommand\eg{e.g.,}								
\newcommand\fday{\hbox{$.\!\!^{\rm d}$}}			
\newcommand\msun{\hbox{$M_{\odot}$}}				
\newcommand\rsun{\hbox{$R_{\odot}$}}				
\newcommand\teff{\hbox{$T_{\rm eff}$}}				
\newcommand\ebv{\hbox{$E_{\rm B-V}$}}				
\newcommand\egr{\hbox{$E_{\rm g-r}$}}				
\newcommand\mg{\hbox{$M_{\rm G}$}}					
\newcommand\av{\hbox{$A_{\rm V}$}}					
\newcommand\ag{\hbox{$A_{\rm G}$}}					
\newcommand\gbp{\hbox{$G_{\rm BP}$}}				
\newcommand\grp{\hbox{$G_{\rm RP}$}}				
\newcommand\rv{\hbox{$R_{\rm V}$}}					

\begin{document}

\title{ZTF J185259.31+124955.2: A new evolved disc-eclipsing \\ binary system}

\author{Klaus Bernhard\inst{\ref{inst:bav}}
	\and Christopher Lloyd\inst{\ref{inst:sussex}}
}

\institute{Bundesdeutsche Arbeitsgemeinschaft f\"{u}r Ver\"{a}nderliche Sterne e.V. (BAV), Munsterdamm 90, D-12169 Berlin \email{klaus.bernhard@liwest.at}\label{inst:bav}
	\and 
Department of Physics and Astronomy, University of Sussex, Brighton, BN1 9QH \email{C.Lloyd@sussex.ac.uk}\label{inst:sussex}
}

\date{Received 1 April 2024 / Accepted 20 May 2024}

\abstract{Discs in long-period eclipsing binary systems are rare and can lead to extraordinary eclipsing events. ZTF J185259.31+124955.2 was identified as a candidate disc-eclipsing system through a continuing search programme of ZTF variables with a near-IR excess in the WISE data. Examination of the combined ZTF and ATLAS photometry shows seven eclipses since 2017 with depths of 0\fm34 in all bands on a period of $289.57\pm0.09$\,d. The eclipse width is $\sim 40$\,d but this and the profile evolve over time. Comparison with library spectra shows that the spectral energy distribution from the available photometry is consistent with an early K-type giant, and fitting black-body profiles suggests $\teff \sim 4000$\,K for the stellar component, with a cool component having $\teff < 500$\,K. The reddening and distance, and hence the luminosity place the star within the giant branch. The most likely scenario is that the system is in a state of rapid evolution following Case B/C mass transfer into an extended disc around an unseen companion.}	

	
\titlerunning{ZTF J185259.31+124955.2: A new disc-eclipsing binary system}
\authorrunning{Bernhard \& Lloyd}

\maketitle

\section{Introduction}

Although very rare, there are several long-period eclipsing binaries that show evidence of large, thin equatorial dust discs. These systems are unusual and complex variables, and frequently show changes in behaviour over time. They are an inhomogeneous group, but divide very broadly into pre-main-sequence and evolved systems. The most well-known eclipsing-disc system is the 27.1-year binary \object{$\epsilon$~Aurigae}, in which the brighter component, apparently an F-type giant, is eclipsed by an accretion disc surrounding an unseen, more massive companion, most likely a B-type star. The system is probably the result of late Case~B or Case~C mass transfer 
\citep[see][]{2018MNRAS.476.5026G}. 
The other well-known system is \object{EE~Cephei}, which has a period of 5.6 years and consists of a Be-star and a dark dusty disc around an unseen companion 
\citep[see][]{2012A&A...544A..53G, 
2020A&A...639A..23P}. 
More recently the OGLE variable, \object{OGLE-LMC-ECL-11893} (OGLE J05172127$-$6900558) has also been identified as another evolved system containing a B9\,IIIe star, which undergoes a complex two-stage, asymmetric eclipse every 468 days, by an extended disc, with a compact core 
\citep[see][]{2014ApJ...788...41D, 
2014ApJ...797....6S}. 

Periodic variations due to dust are also seen in pre-main-sequence stars, but differences in geometry and the distribution of dust leads to a disparate mix of single, binary and triple systems.
Both \object{V1400~Cen} \citep[= J1407, $P\sim5.4$\,yr ?][]{2012AJ....143...72M,2021A&A...652A.117B} 
and \object{V928~Tau} \citep[$P>66$\,d,][]{2020AJ....160..285V} show deep eclipse-like features, but without any clear periodicity. 
The unusual pre-MS stars
\object{V718~Per} \citep[$P=4.7$\,yr][]{2008A&A...489.1233G} 
and CHS~7797 \citep[= \object{V2187~Ori},][]{2012A&A...544A.112R,2013A&A...551A..44R}
both have amplitudes of $\Delta I\sim0\fm7$ and approximately sinusoidal variations, but the periods are very different at 4.7\,yr and 17.8\,d respectively, and V718~Per is probably a single star, while the nature of CHS~7797 is still unclear.
In contrast, the light curves of the low-amplitude Orion-type variables 
\object{VSSG~26} \citep[= WL~4, $P=131$\,d,][]{2008ApJ...684L..37P}
and
\object{YLW~16A} \citep[$P=93$\,d,][]{2013A&A...554A.110P} 
are more nearly square waves.

VSSG~26, YLW~16A and CHS~7797 are often compared to Kearns-Herbst 15D 
\citep[\object{KH 15D} = V582 Mon,][]{1998AJ....116..261K}, 
which now constitutes the prototype of category of circumbinary disc systems. Photometric and spectroscopic observations have shown that the circumbinary disc is tilted relative to the central, highly eccentric binary. 
In a previous study \citet{2022ApJ...933L..21Z} 
reported a search for additional KH~15D objects, and identified two further candidates, 
\object{ZTF J202055.22+381323.1} ('Bernhard 1') and \object{ZTF J071445.39$-$090152.1} ('Bernhard 2'). 
A characteristic of KH 15D, as well as the two other identified candidates, is a large amplitude exceeding 2~magnitudes in the I-band at the time of their discovery. However, long-term time series analysis of KH 15D, such as the study by 
\citet{2021MNRAS.503.1599P} 
reveals that these objects do not consistently display such high amplitude variations.
The aim of this new study was to search for additional disc-related long-period eclipsing binary systems in the ZTF database, focusing on lower amplitude variations. 

\begin{figure*}
	\centering
	\includegraphics[width=17cm]{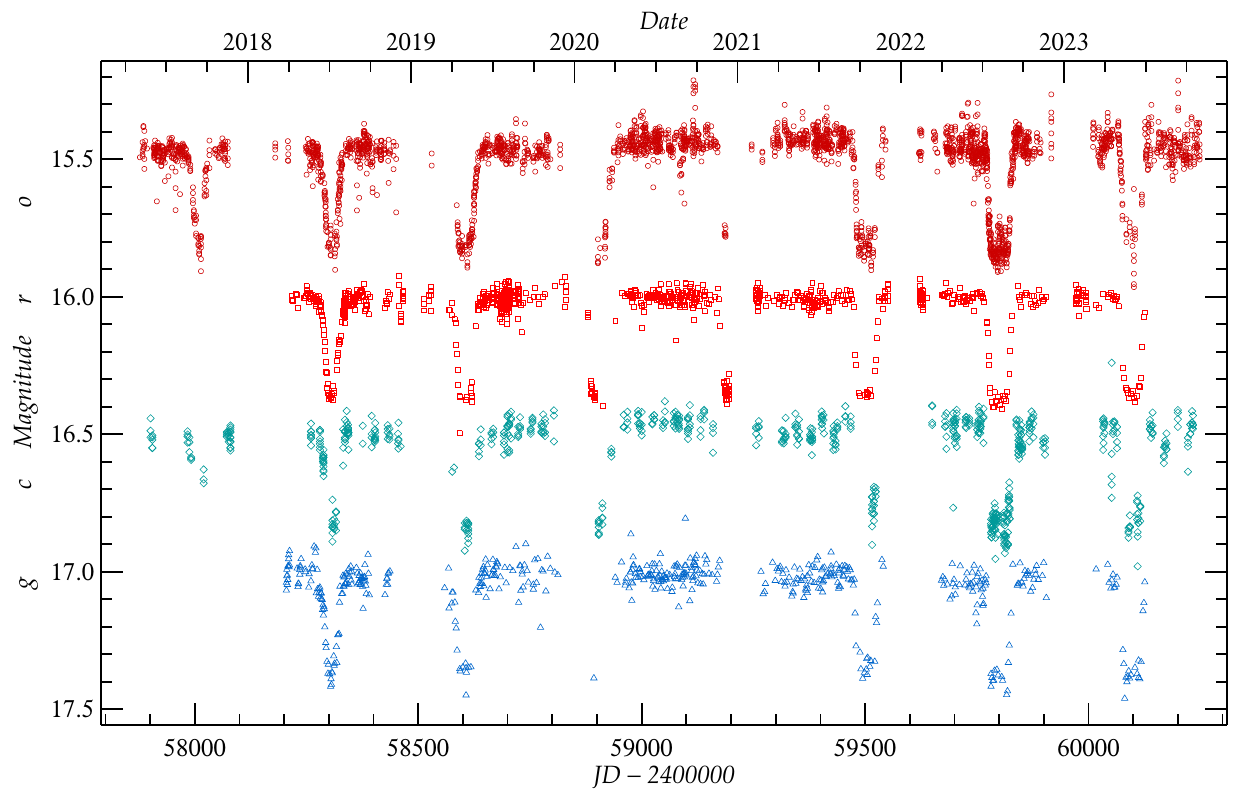}
	\caption{Light curve from ATLAS and ZTF showing the bands with different symbols. These are in order of decreasing effective wavelength, from  \emph{o} (top) through \emph{r} and \emph{c} to \emph{g} (bottom), with small offsets to aid visibility ($r+0.3$, $c-0.1$, $g-0.5$).}
	\label{fig:epoch}
\end{figure*}

\section{Observations}

The starting point of this search is the ZTF catalogue of variable stars by 
\citet{2020ApJS..249...18C}, 
which contains about 780,000 periodic variables with provisional classifications. The catalogue is based on data from the Zwicky Transient Facility (ZTF), which is a time-domain survey using the 48-inch Palomar Schmidt Telescope and has been in operation since 2017. Observations are made in the three different Sloan passbands \emph{g} (414--546\,nm), \emph{r} (566--721\,nm) and partly \emph{i} (721--873\,nm) on a cadance of 1--2\,d, and to a limiting magnitude of 20.5. ZTF data are therefore ideally suited for studies of variable stars 
\citep{2019PASP..131f8003B, 
2019PASP..131a8003M}. 
Supplementary data were acquired through the 'cyan' \emph{c} band (420–650\,nm), and the 'orange' \emph{o} band (560--820\,nm) of the ATLAS Forced Photometry web service
\citep{2018PASP..130f4505T, 
2018AJ....156..241H, 
2021TNSAN...7....1S}. 
The initial study \citep{2022ApJ...933L..21Z} 
focused on identifying eclipsing variable objects with periods greater than 10 days and amplitudes exceeding 1.5 magnitudes in the two ZTF catalogues of variable stars, which included also a list of approximately 1,000,000 suspected variables. By contrast, this research is centred on investigating smaller amplitude variations, specifically in the range of about 0.5 to 1.5 magnitudes. Due to the selection of smaller amplitudes, a substantially larger pool of candidates emerged (19\,649 compared to 1\,041). As an additional selection, only those objects displaying clear indications of a cool disc in the 
WISE colours \citep{2014yCat.2328....0C} 
were examined in detail. This process yielded 19 objects, which were then subjected to visual inspection. 

\section{Results}

Examination of the light curves of the 19 candidates suggests that the majority are, superficially, eclipsing binary systems without any particularly distinguishing features. However, one shows an evolving eclipse, characteristic of a disc system, which is the subject of this paper, while two others show different unusual features and will be reported elsewhere (Bernhard \& Lloyd, in prep.).
The most promising candidate, ZTF J185259.31+124955.2, is categorized as an EA-type variable with a 290-day period by 
\citet{2020ApJS..249...18C}. 
ZTF J185259.31+124955.2 is recorded as \emph{Gaia} DR3 4506139331756845568 (18 52 59.315 +12 49 55.287, J2000), with an average brightness of $G=15.63$ from the \emph{Gaia} DR3 data as listed in VizieR. The Gaia Apsis processing gives a distance of 2900\,pc leading to an absolute magnitude of $\mg = 1.4$, and its effective temperature $\teff = 4831$\,K suggests that the star is a giant
\citep{2023A&A...674A...1G, 
2022yCat.1355....0G}. 

The light curve from 2017 is shown in Fig.~\ref{fig:epoch} for the ATLAS \emph{o} and \emph{c} bands, and the ZTF \emph{g} and \emph{r} bands. A significant number of ZTF $i$-band observations were made during 2020, but these miss the eclipses and provide only the out-of-eclipse magnitude in this band, so are not shown here. A small number of ATLAS observations are also available from 2015 and 2016, but these are sparse and again miss the eclipses, so they are not shown here, although they are included in the analysis later. The different bands are shown close to their natural magnitudes, but with small offsets to aid visibility ($r+0.3$, $c-0.1$, $g-0.5$), and this means that the bands appear in order of decreasing effective wavelength, from \emph{o}, through \emph{r} and \emph{c} to \emph{g}. The ATLAS bands are both very broad, $\sim220$\,nm, and their short-wavelength limits are essentially coincident with those of the ZTF \emph{g} and \emph{r} bands, but these are narrower, so they effectively cover approximately the shorter-wavelength half of the \emph{c} and \emph{o} bands. It should also be pointed out that the out-of-eclipse magnitude in the ATLAS photometry is not completely constant, and shows a small brightening through the data of perhaps 0\fm05, and other small coherent excursions, that are not seen in the ZTF data.

\begin{figure*}
	\centering
	\includegraphics[width=\columnwidth]{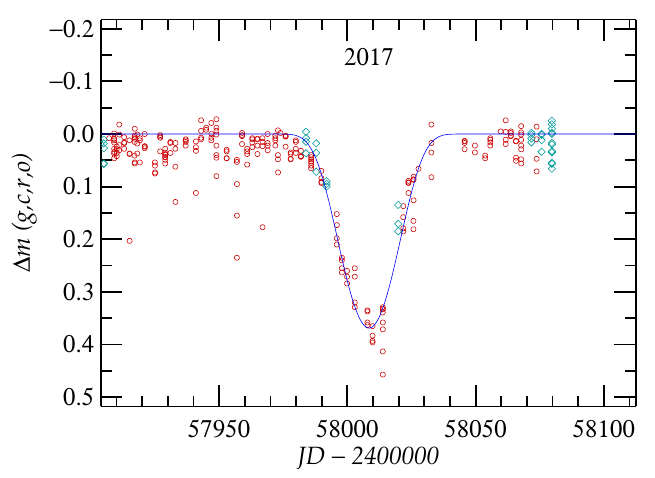}	\includegraphics[width=\columnwidth]{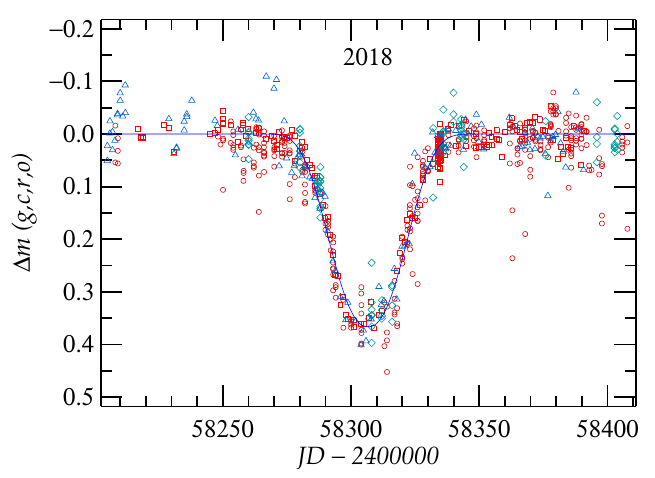}
	\includegraphics[width=\columnwidth]{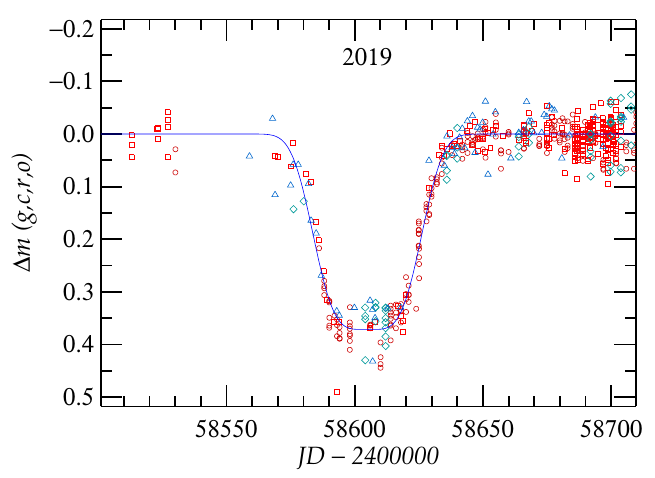}
	\includegraphics[width=\columnwidth]{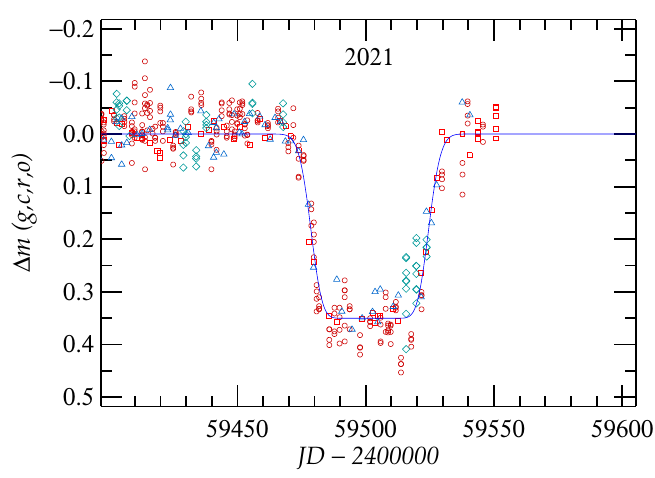}
	\includegraphics[width=\columnwidth]{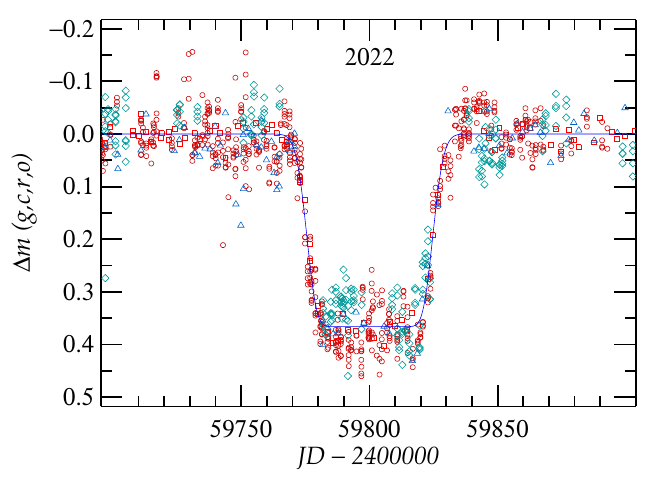}
	\includegraphics[width=\columnwidth]{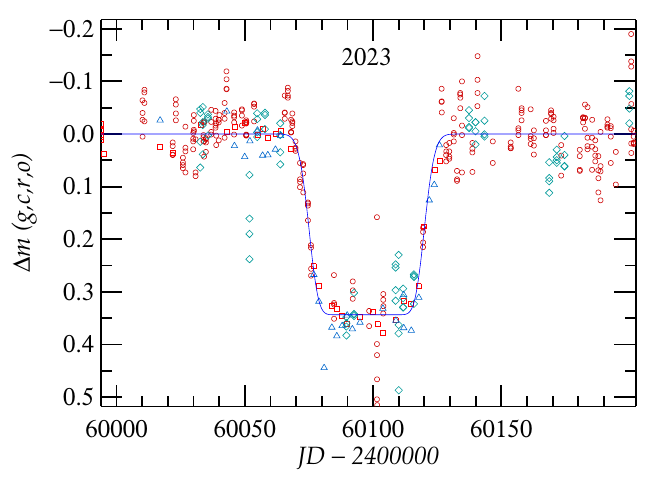}
	\caption{Evolution of the shape and width of the eclipse using the magnitude offsets for all the bands as available, relative to the out-of-eclipse values. The symbols are as before.}
	\label{fig:minima}
\end{figure*}

The light curve itself shows evidence of eight eclipses, with a period close to 290\,d, as suggested by Chen et al., and this is supported by a high-order Fourier fit, which confirms that the variation is periodic. However, even at this point it is clear that the later eclipses are broader than the earlier ones, and also that the eclipse depth is constant across all four bands at approximately 0\fm4.  

In an effort to quantify the properties of the light curves they have been modelled using the methodology described by
\citet{2015A&A...584A...8M}, 
which can fit a wide variety light curve shapes, from exo-planet and simple detached, partial and total EBs to complex W~UMa types, using a small number of parameters. It is not a physical representation, but a mathematical parameterization of the light curve. The expression used is their Equation~13 in the single, primary eclipse form. After some experimentation it was decided to eliminate the effect of the 'curvature' coefficient, $C_k$, by setting this to zero. On the present data sets this term allowed too much flexibility and produced unnatural and unjustified features in the fits. The parameters that can be determined directly from the fits are the central time, the zero level, that is the out-of-eclipse magnitude, and the depth of the minimum. Less directly, it is also possible to measure the full width half maximum (FWHM) of the eclipse profile. Preliminary fits to the individual minima in the different bands immediately revealed that the depths of the eclipses are very similar in all bands, and over time. Using this it was possible to improve the statistics, and coverage, of the individual minima by combining the data from different bands. The two eclipses from 2020 are poorly covered, and the later one is only observed on the ingress, so parameters were derived for seven of the eclipses. The six best-observed minima and their fits are shown in Fig.~\ref{fig:minima}. It is clear that over the period of observations the width of the eclipse has increased, and the shape of the profile has become apparently 'total'. The relevant parameters of the fits are listed in Table~\ref{tab:minima}. 

Examination of the individual minima shows that there is no consistent difference in the depth, or profile in the different bands. The largest difference is probably during the 2021 eclipse where it reaches perhaps 0\fm05 between the blue and the red, but the scatter is large and there are other instances of similar offsets, in both directions, in the out-of-eclipse data, so its significance is not clear. The rms variation of the eclipse depths over time is 0\fm014, and is generally compatible with the measured uncertainties, although the first three eclipses are the deepest, but again the significance, if any, is not clear.

\begin{figure}
	\centering
	\includegraphics[width=\columnwidth]{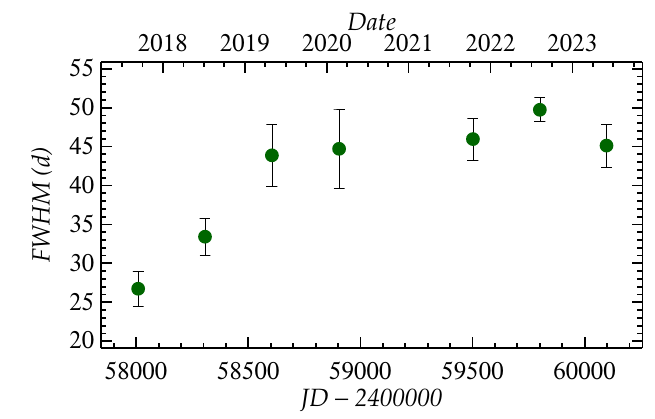}
	\caption{Full width at half minimum of the observed minima from the combined data as given in Table~\ref{tab:minima}.}
	\label{fig:width}
\end{figure}

\begin{table}
	\caption{Parameters of fits to combined minima} \label{tab:minima}
	\vspace{1pt}
	\centering
	\begin{tabular}{crccc}
		\hline \hline
		\addlinespace[2pt] 
		HJD         & O--C  & E  &   Depth & FWHM  \\
		&(d)	&	& (mag) & (d) \\
		\hline
		\addlinespace[2pt] 
		2458008.20(40)  &  $-0.1$   &   0  &  0.375(11)   &  27.2$\pm2.2$  \\
		2458306.65(21)  &  $-0.2$   &   1  &  0.365(9)~   &  33.4$\pm2.4$  \\
		2458605.05(32)  &  $-0.4$   &   2  &  0.370(7)~   &  43.9$\pm3.9$  \\
		2458904.70(50)  &  ~0.7     &   3  &  0.352(9)~   &  44.7$\pm5.0$  \\
		2459501.47(22)  &  ~0.3     &   5  &  0.350(6)~   &  46.4$\pm2.7$  \\
		2459800.16(13)  &  ~0.4     &   6  &  0.365(3)~   &  49.7$\pm1.5$  \\
		2460097.51(25)  &  $-0.8$   &   7  &  0.343(6)~   &  45.1$\pm2.8$  \\
		\hline
	\end{tabular}
\end{table}

However, the variation in the eclipse width is clearly established. It is obvious in the plots of the individual minima and is shown directly in Fig.~\ref{fig:width}. The width increased from 27 to 44\,d in the first three cycles and then appears to have remained constant since then. However, the profile continued to evolve from 2019 to 2021 with ingress and egress becoming progressively steeper, but with little change in the FWHM. The change in width of the early eclipses is not matched by a change in gradient of the profile. and it appears that the eclipse simply broadened until 2019, when it becomes total.

\begin{figure}
	\centering
	\resizebox{\hsize}{!}{\includegraphics{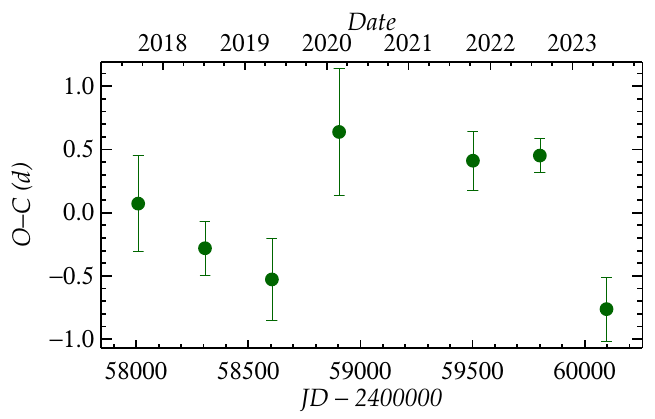}}
	\caption{O--C diagram constructed using the ephemeris in Equation~\ref{eqn:ephem} derived from the timings given in Table~\ref{tab:minima}.}
	\label{fig:o-c}
\end{figure}

Using the central time of the eclipse it is possible to derive a more reliable value for the period, and an unweighted linear fit gives an ephemeris of
\begin{equation}
	HJD_{\rm MinI} = 2458008.30\pm0.37 + 298.57\pm0.09 \times E \label{eqn:ephem}
\end{equation}
where the residuals are shown in the O--C diagram in Fig.~\ref{fig:o-c}. The rms residual of 0\fday5 and the formal error on the period suggest that the period is very tightly constrained, despite the changes in the shape and width of the eclipse profile, and this suggests that the eclipse process is symmetric with respect to some dynamical clock in the system.

\begin{figure*}
	\centering
	\includegraphics[width=17cm]{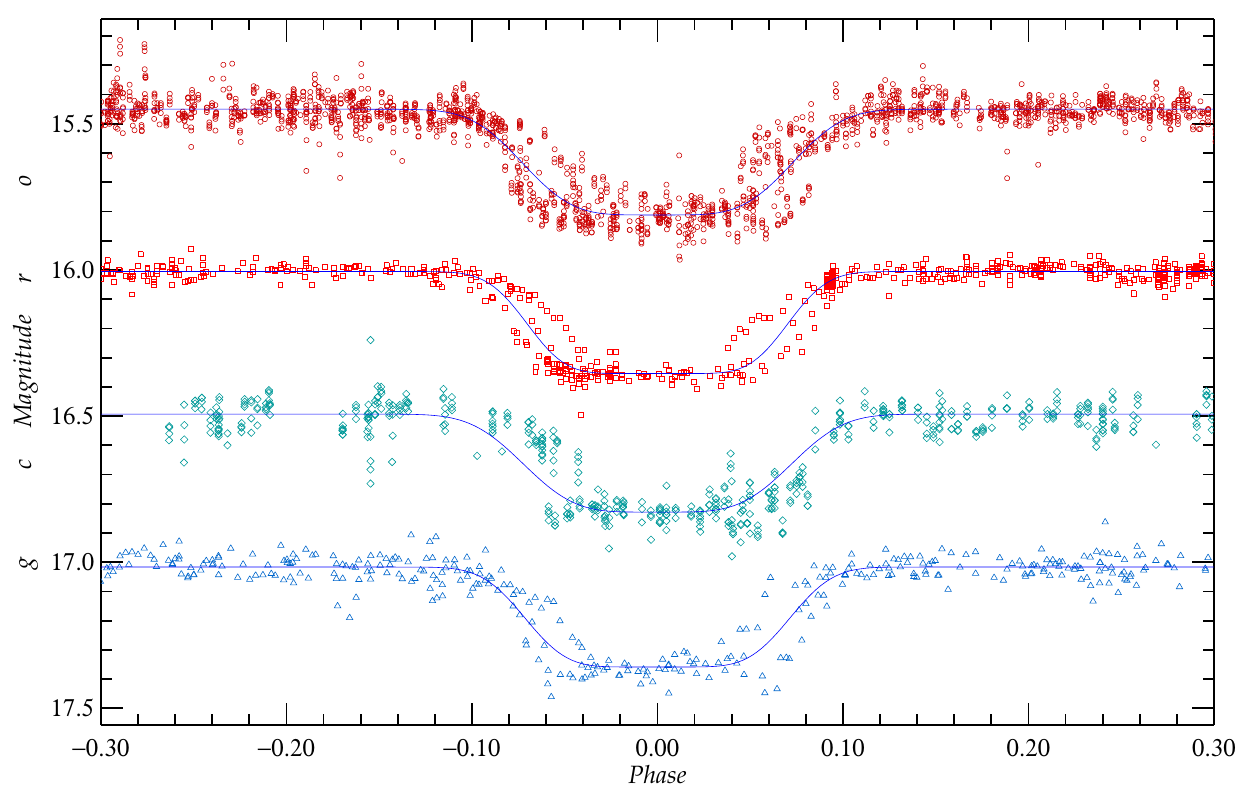}
	\caption{Phased light curve of the ATLAS \emph{o} and \emph{c}-band and ZTF \emph{r} and \emph{g}-band data, based on the ephemeris Equation~\ref{eqn:ephem}. The magnitude offsets and symbols are the same as Fig.~\ref{fig:epoch}.}
	\label{fig:folded}
\end{figure*}

With an improved ephemeris it is possible to phase the individual bands and examine their mean light curves. These are shown in Fig.~\ref{fig:folded} with their individual fits. As the eclipse profile is known to change these fits provide only two useful parameters, the mean out-of-eclipse magnitude and the mean depth of the eclipse for each band. These values are collected in Table~\ref{tab:photometry}. However, the plot also demonstrates that the changes to the eclipses are symmetric and further supports the conclusions from the O--C diagram. Given the constancy of the  ZTF \emph{r}-band  out-of-eclipse photometry, it is probably the most reliable, so the eclipse 
also has constant brightness during minimum light.

\begin{table*}
	\caption{Mean photometric data} \label{tab:photometry}
	\vspace{1pt}
	\centering
	\begin{tabular}{lccccc}
		\hline \hline
		\addlinespace[2pt] 
		& ZTF \emph{g} & ATLAS \emph{c} & ZTF \emph{r} & ATLAS \emph{o} & ZTF \emph{i} \\
		\hline
		\addlinespace[2pt] 
		Maximum & 17.516(2)    & 16.595(2)      &  15.705(2)   & 15.450(1)      &  14.927(3) \\
		Depth   & 0.342(8)     &  0.334(6)      &  0.349(7)    &  0.361(4)      &   --\\
		\hline
	\end{tabular}
\end{table*}

\section{Spectral energy distribution and luminosity}

\begin{figure}
	\centering
	\includegraphics[width=\columnwidth]{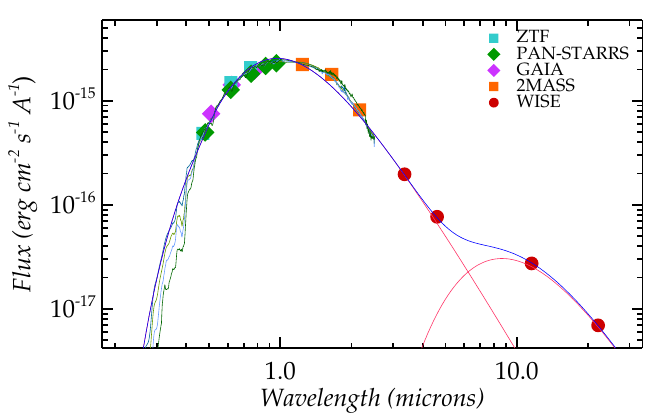}
	\caption{Spectral energy distribution showing the observed fluxes with the optimally reddened G0\,III, G5\,III, K0\,III and K5\,III library spectra. The red lines are the individual black-body fits and the blue line is the combined two-component fit.}
	\label{fig:sed}
\end{figure}

As mentioned earlier the star is most likely a cool giant, and was also selected on the basis of an excess in the WISE infra-red flux. The full spectral energy distribution (SED) has been explored using mean magnitude measurements covering a wide wavelength range from \emph{Gaia} DR3 \emph{G}, \gbp, \grp, 
ZTF \emph{gri}, 
Pan-STARRS \emph{grizy} \citep{2016arXiv161205560C}, 
2MASS $JHK_{\rm s}$ \citep{2003yCat.2246....0C,2006AJ....131.1163S} 
and WISE $W1$, $W2$, $W3$, and $W4$ 
\citep{2010AJ....140.1868W,2014yCat.2328....0C}. 
These have compared to a range of spectra from the 
\cite{1998PASP..110..863P} 
Stellar Flux Library using a minimization scheme that can treat the reddening as a free parameter. 
The photometry was compared with the luminosity class III library spectra of F5, G0, G5, K0, K5, M0 and M3 stars, out to 2\,microns. The best fit was achieved for each of the library spectra with reddening \ebv\ of 1.27, 1.06, 0.96, 0.89, 0.44, 0.32, and 0.06 respectively.  
The best fits were achieved with the G0\,III, G5\,III, K0\,III and K5\,III spectra, which are all very similar and were significantly better than F5\,III and the later spectral types. The observed SED and the optimally reddened fits for these spectra are shown in Fig.~\ref{fig:sed}. While the WISE W1 and W2 photometry at 3.35 and 4.6\,$\mu$m may be consistent with these spectra the W3 and W4 bands beyond 10\,$\mu$m are clearly incompatible and suggest an additional cool component. So, the data have been fitted with a two-component black-body, however, the solution is poorly constrained so some limitations have been imposed. Equal relative errors have been assumed on the photometry, and the reddening has been fixed, in the case illustrated at $\ebv=0.6$. The optimal black-body temperatures are $T_1=3940\pm280$\,K and $T_2=340\pm140$\,K, and with larger reddening, higher temperatures result, with a sensible limit of $T_1\sim 4500$\,K, and it is difficult to force $T_2$ above 400--500\,K. The \emph{Gaia} Apsis processing chain gives $\teff=4830$\,K.

The luminosity of the system is obviously dependent on the distance, and to some extent so is the reddening. However, there is no consensus on the distance. The parallax from \emph{Gaia} is $\Pi=0.153\pm0.039$\,mas, which leads directly to a rather uncertain $d=6500^{+2200}_{-1300}$\,pc. 
\citet{2021AJ....161..147B} 
provide two Gaia-derived distances, one from geometric distance posterior (\emph{rgeo}) of $5400^{+1700}_{-900}$\,pc, and a second from the photogeometric distance posterior (\emph{rpgeo}) of $4740^{+925}_{-670}$\,pc, which lead to a 1-sigma range of 4100--7100\,pc. As mentioned earlier the \emph{Gaia} Apsis processing gives a surprisingly low distance of 2609\,pc, and well outside the range of the other estimates.

\begin{figure}
	\centering
	\includegraphics[width=\columnwidth]{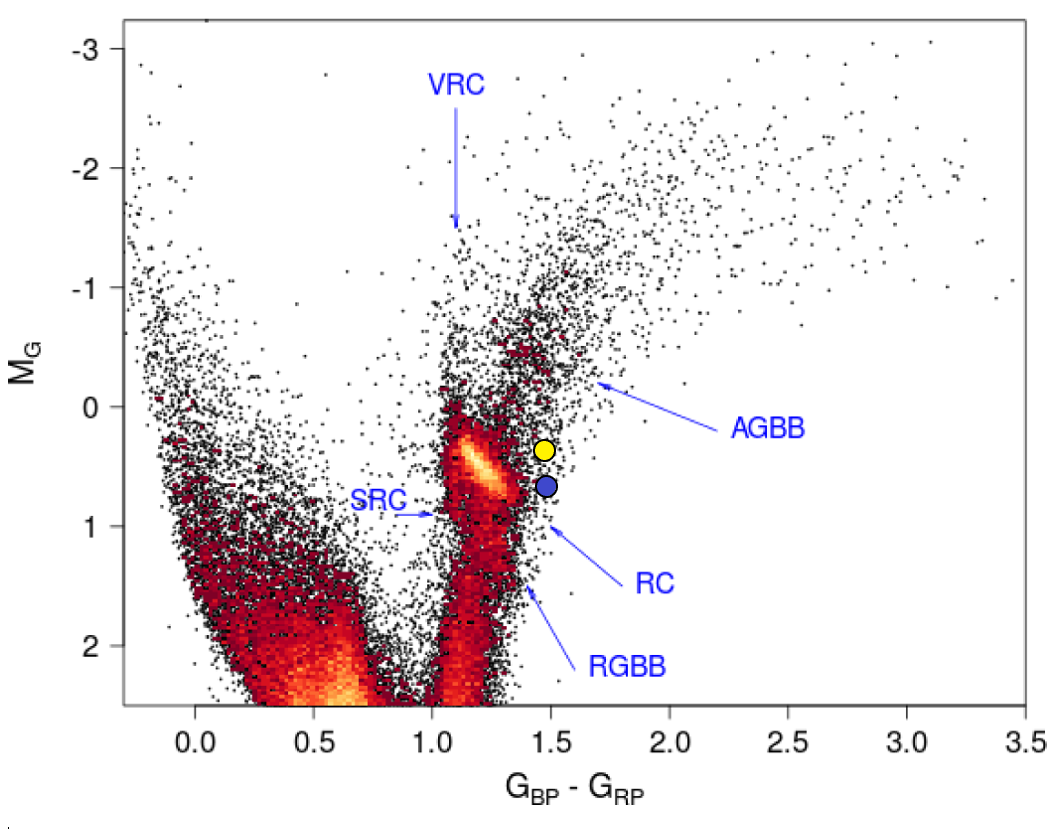}
	\caption{Position of the star in the HR diagram. \mg\ is calculated using the minimum luminosity and assuming $\ebv=0.6$ for the \emph{Gaia} \emph{rgeo} (blue) and \emph{rpgeo} (yellow) distances. The points are superimposed on the \emph{Gaia} data for low-extinction nearby giants \citep[Fig. 10 in][]{2018A&A...616A..10G} and lie close to the Red Clump, and well within the distribution of a larger sample of stars (see \eg\ Fig. 5 in the same paper).}
	\label{fig:hr}
\end{figure}

Based on the \emph{Gaia} photometry Apsis gives the visual extinction as $\av=3.07$, which with $\rv=3.1$ leads to $\ebv=0.99$.
An independent measure of the reddening is provided by the 3D Dust Maps of 
\citet{2019ApJ...887...93G} 
who find $\egr=0.60^{+0.08}_{-0.01}$ at 4740\,pc, and that the reddening plateaus at $\egr=0.69$ beyond 6000\,pc.
Unfortunately \citet{2018A&A...616A.132L} 
do not provide any reddening estimates at this distance. From the available data, and assuming $\egr=\ebv$ at this level of precision, then the system has  $\ebv>0.6$, but probably $<1.0$, and this is broadly consistent with the findings from the SED. With the  reddening in this range then $\av\sim2$--3 magnitudes, and using the
\citet{2019ApJ...877..116W} relationship, $\ag=0.789\av$, then the extinction in \emph{G}, $\ag\sim1.6$--2.4.  
With a mean magnitude out-of-eclipse of $G=15.63$ then for the \emph{rgeo} distance at 4740\,pc $\mg=0.7$ to $-0.1$, while for the \emph{rpgeo} distance at 5400\,pc,  
$\mg=0.4$ to $-0.4$. The mean observed out-of-eclipse $\gbp-\grp=2.30$ so with $\ebv=0.6$ and using the relationship $\ebv=0.76E_{(\gbp-\grp)}$ then $(\gbp-\grp)_0=1.50$. The dereddened \mg\ and $(\gbp-\grp)_0$ magnitudes are plotted for both cases in the HR diagram in Fig.~\ref{fig:hr}, and place the star well within the giant branch.

\section{Discussion}

As the object presents as a K-type giant, both in terms of luminosity and its SED, that immediately suggests that it is not a young or pre-main-sequence system, and must be evolved, like $\epsilon$~Aur or EE~Cep, but on a very much smaller scale. It is most likely the result of late Case~B or Case~C mass transfer, where the K-type star has lost a high proportion of its mass and is now the secondary in a system, where the new primary is obscured by the accreted material, some of which is formed into an extended disc. From the SED it is clear that there is a faint, cool component in the system, and this is most likely emission from dust, which is also responsible for the grey eclipses. There is no indication of any other variation during the cycle, and nothing that could be interpreted as a secondary eclipse. 

The eclipse is 0\fm34 deep so this means that at minimum some 73\% of the light in the system is uneclipsed. It is possible that the star is fully occulted, but that the optical depth is sufficiently low that a only a small fraction of light is scattered. However, if the disc is relatively transparent then the other component should be visible, so this argues against a thin equatorial disc in the line of sight. If the disc has significant thickness, or is inclined to the orbital plane, then it is possible that the face of the K~star is not completely covered, perhaps by a thin ring or other moving features in, or on the boundary of the disc. At present there is very little information to constrain the geometry of the system, but the eclipse variation demands some asymmetry, either as an inclined or thick disc. Another obvious question is what limits the duration of the eclipse, while simultaneously limiting its depth in a way that it appears to reach a saturation limit. One possibility is that eclipsed star contributes only 30\% of the light to the system and is completely obscured by a variable disc, but that would require that the most visible component, the K~giant, is not involved, and would effectively be a third light.

\begin{figure}
	\centering
	\includegraphics[width=\columnwidth]{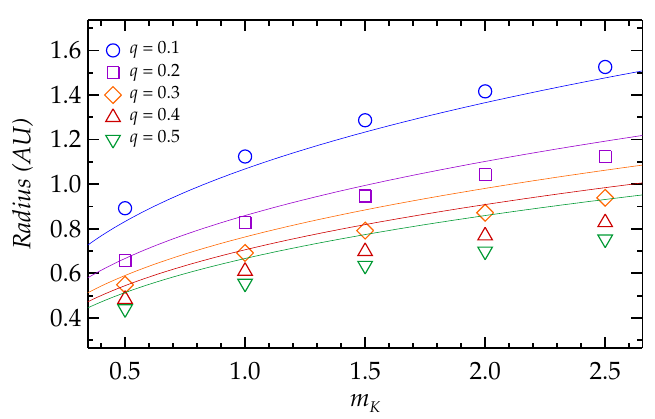}
	\caption{Radius of the Roche lobe around the unseen component. The radius is calculated for sample mass ratios, \emph{q}, as given by the symbols, for five possible masses of the K~giant, assuming $r_K=16$\,\rsun\ (see text). The solid lines are the radii of an equatorial disc corresponding to the phase coverage of the eclipse, $\Delta\phi=0.1$, colour-coded for the values of \emph{q}. For $q=0.1$ the equatorial disc lies within the Roche lobe, but for increasing values of $q$ the radius of the eclipsing material progressively exceeds the Roche lobe by up to 22\% at $q=0.5$.}
	\label{fig:radius}
\end{figure}

Assuming the least contrived arrangement of the components, that is with the lower-mass K-star secondary and a more massive dark primary component, with its circumstellar disc, then it is possible to draw some preliminary conclusions. 
Taking a range of K-star masses, $m_K=0.5...2.5$\,\msun\ \citep{2011MNRAS.414.2594H}, and mass ratios $q=m_K/m_{dark}=0.1...0.5$, with the known period, it is possible to calculate the size of the system, 
which ranges from $a=1.5$ to 2.6\,AU for $q=0.1$, to $a=1.0$ to 1.7\,AU for $q=0.5$.
Using the
\citet{1983ApJ...268..368E} approximation, it is possible to calculate the Roche lobe radii of both components for the various combinations of $m_K$ and $q$.
The results are shown in Fig.~\ref{fig:radius} where the symbols show the Roche lobe radii of the dark component.   Assuming that the eclipse is due to an equatorial disc around the dark component then its radius can be calculated from the phase coverage of the eclipse, taken as $\Delta\phi=0.1$, and these are shown as the solid lines. 
For $q=0.1$ the radius of the eclipsing material lies just within the appropriate Roche lobe, but for the larger values of $q$ the radius of the disc exceeds the Roche lobe by 5\% at $q=0.2$ and 22\% at $q=0.5$. If this scenario is physically meaningful then it suggests that the mass ratio $q > 0.2$. The radius of the K~giant is not known, and given the likely history of the system may not be nominal for the spectral type, however, the luminosity and temperature are consistent with what is expected, so the radius probably is as well. Mean radii of K~giants are given as $16\pm2$ and $31\pm6$\,\rsun\ at K1\,III and K4\,III respectively by \citet{1996AJ....111.1705D},
and more broadly red giants have a median radius of 10\,\rsun, 
but these are dominated by red-clump stars \citep{2011MNRAS.414.2594H}. The radius of the K~giant also has a small impact on the calculation of the disc radius given above. There $r_K$ was taken as 16\,\rsun\ but if the value of 31\,\rsun\ is used then the radius of the equatorial disc is reduced generally by $\sim 5$\%, and becomes entirely compatible with $q=0.2$.

For the range of parameters used here the minimum Roche-lobe radius of the K~giant is 68\,\rsun\ (0.32\,AU) for $q=0.1$ rising to 117\,\rsun\ (0.54\,AU) for $q=0.5$, largely irrespective of the total mass, so is significantly larger than the star in all cases. As the mass transfer has now ceased, the Roche lobe radius provides a fossilized record of the radius of the K~giant at the time the system became detached. While all the possible radii are consistent with large giants or supergiants, the smallest radius for $q=0.1$ represents the lowest fractional change, and may offer additional weak support for that value.

\section{Conclusions}

ZTF J185259.31+124955.2 is shown to be an unusual long-period eclipsing binary with precise periodicity (Equation~\ref{eqn:ephem}), however the eclipse width and profile have changed symmetrically during the observations over the past seven years. The depth of the eclipse appears to have remained constant over time (see Fig.~\ref{fig:minima} and Table~\ref{tab:minima}), and with wavelength (see Fig.~\ref{fig:folded} and Table~\ref{tab:photometry}), so the eclipse is grey, meaning that the occulting material is optically thin dust. The optical-NIR SED is consistent with an early K-type giant and with the observed reddening $\ebv\approx0.6$, and also shows a cool component consistent with extended dust. Any model of the system will need to explain the apparent limit or saturation of both the eclipse depth, which appears to be constant despite the other changes, and in width, where the eclipse widens and then apparently steepens. These changes appear to require some evolution or asymmetry in the disc. Assuming the most obvious scenario, the width of the eclipse is consistent with a disc lying within the Roche radius of the dark component for $q < 0.2$, as demonstrated in Fig.~\ref{fig:radius}. The most obvious requirement for new observations is a spectroscopic study of the system, to further understand what is being eclipsed, and for radial velocities to place some constraints on the masses involved.

\begin{acknowledgements}
This paper made use of data from VizieR catalog access tool, of the interactive sky atlas Aladin, CDS, Strasbourg, France, the NASA/IPAC Infrared Science Archive and of the International Variable Star Index (VSX) database of the American Association of Variable Star Observers (AAVSO, Cambridge, Massachusetts, USA). It also utilizes data from the Asteroid Terrestrial-impact Last Alert System (ATLAS) project. This work has also made use of data from the European Space Agency (ESA) mission \emph{Gaia} (https://www.cosmos.esa.int/gaia).
\end{acknowledgements}

%
%

\begin{thebibliography}{39}
	\expandafter\ifx\csname natexlab\endcsname\relax\def\natexlab#1{#1}\fi
	
	\bibitem[{{Bailer-Jones} {et~al.}(2021){Bailer-Jones}, {Rybizki}, {Fouesneau},
		{Demleitner}, \& {Andrae}}]{2021AJ....161..147B}
	{Bailer-Jones}, C.~A.~L., {Rybizki}, J., {Fouesneau}, M., {Demleitner}, M., \&
	{Andrae}, R. 2021, \aj, 161, 147
	
	\bibitem[{{Barmentloo} {et~al.}(2021){Barmentloo}, {Dik}, {Kenworthy},
		{Mamajek}, {Hambsch}, {Reichart}, {Rodriguez}, \& {van
			Dam}}]{2021A&A...652A.117B}
	{Barmentloo}, S., {Dik}, C., {Kenworthy}, M.~A., {et~al.} 2021, \aap, 652, A117
	
	\bibitem[{{Bellm} {et~al.}(2019){Bellm}, {Kulkarni}, {Barlow}, {Feindt},
		{Graham}, {Goobar}, {Kupfer}, {Ngeow}, {Nugent}, {Ofek}, {Prince}, {Riddle},
		{Walters}, \& {Ye}}]{2019PASP..131f8003B}
	{Bellm}, E.~C., {Kulkarni}, S.~R., {Barlow}, T., {et~al.} 2019, \pasp, 131,
	068003
	
	\bibitem[{{Chambers} {et~al.}(2016){Chambers}, {Magnier}, {Metcalfe},
		{Flewelling}, {Huber}, {Waters}, {Denneau}, {Draper}, {Farrow}, {Finkbeiner},
		{Holmberg}, {Koppenhoefer}, {Price}, {Rest}, {Saglia}, {Schlafly}, {Smartt},
		{Sweeney}, {Wainscoat}, {Burgett}, {Chastel}, {Grav}, {Heasley}, {Hodapp},
		{Jedicke}, {Kaiser}, {Kudritzki}, {Luppino}, {Lupton}, {Monet}, {Morgan},
		{Onaka}, {Shiao}, {Stubbs}, {Tonry}, {White}, {Ba{\~n}ados}, {Bell},
		{Bender}, {Bernard}, {Boegner}, {Boffi}, {Botticella}, {Calamida},
		{Casertano}, {Chen}, {Chen}, {Cole}, {Deacon}, {Frenk}, {Fitzsimmons},
		{Gezari}, {Gibbs}, {Goessl}, {Goggia}, {Gourgue}, {Goldman}, {Grant},
		{Grebel}, {Hambly}, {Hasinger}, {Heavens}, {Heckman}, {Henderson}, {Henning},
		{Holman}, {Hopp}, {Ip}, {Isani}, {Jackson}, {Keyes}, {Koekemoer}, {Kotak},
		{Le}, {Liska}, {Long}, {Lucey}, {Liu}, {Martin}, {Masci}, {McLean}, {Mindel},
		{Misra}, {Morganson}, {Murphy}, {Obaika}, {Narayan}, {Nieto-Santisteban},
		{Norberg}, {Peacock}, {Pier}, {Postman}, {Primak}, {Rae}, {Rai}, {Riess},
		{Riffeser}, {Rix}, {R{\"o}ser}, {Russel}, {Rutz}, {Schilbach}, {Schultz},
		{Scolnic}, {Strolger}, {Szalay}, {Seitz}, {Small}, {Smith}, {Soderblom},
		{Taylor}, {Thomson}, {Taylor}, {Thakar}, {Thiel}, {Thilker}, {Unger},
		{Urata}, {Valenti}, {Wagner}, {Walder}, {Walter}, {Watters}, {Werner},
		{Wood-Vasey}, \& {Wyse}}]{2016arXiv161205560C}
	{Chambers}, K.~C., {Magnier}, E.~A., {Metcalfe}, N., {et~al.} 2016, arXiv
	e-prints, arXiv:1612.05560
	
	\bibitem[{{Chen} {et~al.}(2020){Chen}, {Wang}, {Deng}, {de Grijs}, {Yang}, \&
		{Tian}}]{2020ApJS..249...18C}
	{Chen}, X., {Wang}, S., {Deng}, L., {et~al.} 2020, \apjs, 249, 18
	
	\bibitem[{{Cutri} {et~al.}(2003){Cutri}, {Skrutskie}, {van Dyk}, {Beichman},
		{Carpenter}, {Chester}, {Cambresy}, {Evans}, {Fowler}, {Gizis}, {Howard},
		{Huchra}, {Jarrett}, {Kopan}, {Kirkpatrick}, {Light}, {Marsh}, {McCallon},
		{Schneider}, {Stiening}, {Sykes}, {Weinberg}, {Wheaton}, {Wheelock}, \&
		{Zacarias}}]{2003yCat.2246....0C}
	{Cutri}, R.~M., {Skrutskie}, M.~F., {van Dyk}, S., {et~al.} 2003, VizieR Online
	Data Catalog, II/246
	
	\bibitem[{{Cutri} {et~al.}(2021){Cutri}, {Wright}, {Conrow}, {Fowler},
		{Eisenhardt}, {Grillmair}, {Kirkpatrick}, {Masci}, {McCallon}, {Wheelock},
		{Fajardo-Acosta}, {Yan}, {Benford}, {Harbut}, {Jarrett}, {Lake}, {Leisawitz},
		{Ressler}, {Stanford}, {Tsai}, {Liu}, {Helou}, {Mainzer}, {Gettngs},
		{Gonzalez}, {Hoffman}, {Marsh}, {Padgett}, {Skrutskie}, {Beck}, {Papin}, \&
		{Wittman}}]{2014yCat.2328....0C}
	{Cutri}, R.~M., {Wright}, E.~L., {Conrow}, T., {et~al.} 2021, VizieR Online
	Data Catalog, II/328
	
	\bibitem[{{Dong} {et~al.}(2014){Dong}, {Katz}, {Prieto}, {Udalski},
		{Kozlowski}, {Street}, {Bramich}, {Tsapras}, {Hundertmark}, {Snodgrass},
		{Horne}, {Dominik}, \& {Figuera Jaimes}}]{2014ApJ...788...41D}
	{Dong}, S., {Katz}, B., {Prieto}, J.~L., {et~al.} 2014, \apj, 788, 41
	
	\bibitem[{{Dyck} {et~al.}(1996){Dyck}, {Benson}, {van Belle}, \&
		{Ridgway}}]{1996AJ....111.1705D}
	{Dyck}, H.~M., {Benson}, J.~A., {van Belle}, G.~T., \& {Ridgway}, S.~T. 1996,
	\aj, 111, 1705
	
	\bibitem[{{Eggleton}(1983)}]{1983ApJ...268..368E}
	{Eggleton}, P.~P. 1983, \apj, 268, 368
	
	\bibitem[{{Gaia Collaboration}(2022)}]{2022yCat.1355....0G}
	{Gaia Collaboration}. 2022, VizieR Online Data Catalog, I/355
	
	\bibitem[{{Gaia Collaboration} {et~al.}(2018){Gaia Collaboration}, {Babusiaux},
		{van Leeuwen}, {Barstow}, {Jordi}, {Vallenari}, {Bossini}, {Bressan},
		{Cantat-Gaudin}, {van Leeuwen}, {Brown}, {Prusti}, {de Bruijne},
		{Bailer-Jones}, {Biermann}, {Evans}, {Eyer}, {Jansen}, {Klioner}, {Lammers},
		{Lindegren}, {Luri}, {Mignard}, {Panem}, {Pourbaix}, {Randich}, {Sartoretti},
		{Siddiqui}, {Soubiran}, {Walton}, {Arenou}, {Bastian}, {Cropper}, {Drimmel},
		{Katz}, {Lattanzi}, {Bakker}, {Cacciari}, {Casta{\~n}eda}, {Chaoul}, {Cheek},
		{De Angeli}, {Fabricius}, {Guerra}, {Holl}, {Masana}, {Messineo}, {Mowlavi},
		{Nienartowicz}, {Panuzzo}, {Portell}, {Riello}, {Seabroke}, {Tanga},
		{Th{\'e}venin}, {Gracia-Abril}, {Comoretto}, {Garcia-Reinaldos}, {Teyssier},
		{Altmann}, {Andrae}, {Audard}, {Bellas-Velidis}, {Benson}, {Berthier},
		{Blomme}, {Burgess}, {Busso}, {Carry}, {Cellino}, {Clementini}, {Clotet},
		{Creevey}, {Davidson}, {De Ridder}, {Delchambre}, {Dell'Oro}, {Ducourant},
		{Fern{\'a}ndez-Hern{\'a}ndez}, {Fouesneau}, {Fr{\'e}mat}, {Galluccio},
		{Garc{\'\i}a-Torres}, {Gonz{\'a}lez-N{\'u}{\~n}ez}, {Gonz{\'a}lez-Vidal},
		{Gosset}, {Guy}, {Halbwachs}, {Hambly}, {Harrison}, {Hern{\'a}ndez},
		{Hestroffer}, {Hodgkin}, {Hutton}, {Jasniewicz}, {Jean-Antoine-Piccolo},
		{Jordan}, {Korn}, {Krone-Martins}, {Lanzafame}, {Lebzelter}, {L{\"o}ffler},
		{Manteiga}, {Marrese}, {Mart{\'\i}n-Fleitas}, {Moitinho}, {Mora}, {Muinonen},
		{Osinde}, {Pancino}, {Pauwels}, {Petit}, {Recio-Blanco}, {Richards},
		{Rimoldini}, {Robin}, {Sarro}, {Siopis}, {Smith}, {Sozzetti}, {S{\"u}veges},
		{Torra}, {van Reeven}, {Abbas}, {Abreu Aramburu}, {Accart}, {Aerts},
		{Altavilla}, {{\'A}lvarez}, {Alvarez}, {Alves}, {Anderson}, {Andrei},
		{Anglada Varela}, {Antiche}, {Antoja}, {Arcay}, {Astraatmadja}, {Bach},
		{Baker}, {Balaguer-N{\'u}{\~n}ez}, {Balm}, {Barache}, {Barata}, {Barbato},
		{Barblan}, {Barklem}, {Barrado}, {Barros}, {Bartholom{\'e} Mu{\~n}oz},
		{Bassilana}, {Becciani}, {Bellazzini}, {Berihuete}, {Bertone}, {Bianchi},
		{Bienaym{\'e}}, {Blanco-Cuaresma}, {Boch}, {Boeche}, {Bombrun}, {Borrachero},
		{Bouquillon}, {Bourda}, {Bragaglia}, {Bramante}, {Breddels}, {Brouillet},
		{Br{\"u}semeister}, {Brugaletta}, {Bucciarelli}, {Burlacu}, {Busonero},
		{Butkevich}, {Buzzi}, {Caffau}, {Cancelliere}, {Cannizzaro}, {Carballo},
		{Carlucci}, {Carrasco}, {Casamiquela}, {Castellani}, {Castro-Ginard},
		{Charlot}, {Chemin}, {Chiavassa}, {Cocozza}, {Costigan}, {Cowell}, {Crifo},
		{Crosta}, {Crowley}, {Cuypers}, {Dafonte}, {Damerdji}, {Dapergolas}, {David},
		{David}, {de Laverny}, {De Luise}, {De March}, {de Martino}, {de Souza}, {de
			Torres}, {Debosscher}, {del Pozo}, {Delbo}, {Delgado}, {Delgado}, {Diakite},
		{Diener}, {Distefano}, {Dolding}, {Drazinos}, {Dur{\'a}n}, {Edvardsson},
		{Enke}, {Eriksson}, {Esquej}, {Eynard Bontemps}, {Fabre}, {Fabrizio},
		{Faigler}, {Falc{\~a}o}, {Farr{\`a}s Casas}, {Federici}, {Fedorets},
		{Fernique}, {Figueras}, {Filippi}, {Findeisen}, {Fonti}, {Fraile}, {Fraser},
		{Fr{\'e}zouls}, {Gai}, {Galleti}, {Garabato}, {Garc{\'\i}a-Sedano},
		{Garofalo}, {Garralda}, {Gavel}, {Gavras}, {Gerssen}, {Geyer}, {Giacobbe},
		{Gilmore}, {Girona}, {Giuffrida}, {Glass}, {Gomes}, {Granvik}, {Gueguen},
		{Guerrier}, {Guiraud}, {Guti{\'e}}, {Haigron}, {Hatzidimitriou}, {Hauser},
		{Haywood}, {Heiter}, {Helmi}, {Heu}, {Hilger}, {Hobbs}, {Hofmann}, {Holland},
		{Huckle}, {Hypki}, {Icardi}, {Jan{\ss}en}, {Jevardat de Fombelle}, {Jonker},
		{Juh{\'a}sz}, {Julbe}, {Karampelas}, {Kewley}, {Klar}, {Kochoska}, {Kohley},
		{Kolenberg}, {Kontizas}, {Kontizas}, {Koposov}, {Kordopatis},
		{Kostrzewa-Rutkowska}, {Koubsky}, {Lambert}, {Lanza}, {Lasne}, {Lavigne}, {Le
			Fustec}, {Le Poncin-Lafitte}, {Lebreton}, {Leccia}, {Leclerc},
		{Lecoeur-Taibi}, {Lenhardt}, {Leroux}, {Liao}, {Licata}, {Lindstr{\o}m},
		{Lister}, {Livanou}, {Lobel}, {L{\'o}pez}, {Managau}, {Mann}, {Mantelet},
		{Marchal}, {Marchant}, {Marconi}, {Marinoni}, {Marschalk{\'o}}, {Marshall},
		{Martino}, {Marton}, {Mary}, {Massari}, {Matijevi{\v{c}}}, {Mazeh},
		{McMillan}, {Messina}, {Michalik}, {Millar}, {Molina}, {Molinaro},
		{Moln{\'a}r}, {Montegriffo}, {Mor}, {Morbidelli}, {Morel}, {Morris},
		{Mulone}, {Muraveva}, {Musella}, {Nelemans}, {Nicastro}, {Noval},
		{O'Mullane}, {Ord{\'e}novic}, {Ord{\'o}{\~n}ez-Blanco}, {Osborne}, {Pagani},
		{Pagano}, {Pailler}, {Palacin}, {Palaversa}, {Panahi}, {Pawlak},
		{Piersimoni}, {Pineau}, {Plachy}, {Plum}, {Poggio}, {Poujoulet},
		{Pr{\v{s}}a}, {Pulone}, {Racero}, {Ragaini}, {Rambaux}, {Ramos-Lerate},
		{Regibo}, {Reyl{\'e}}, {Riclet}, {Ripepi}, {Riva}, {Rivard}, {Rixon},
		{Roegiers}, {Roelens}, {Romero-G{\'o}mez}, {Rowell}, {Royer}, {Ruiz-Dern},
		{Sadowski}, {Sagrist{\`a} Sell{\'e}s}, {Sahlmann}, {Salgado}, {Salguero},
		{Sanna}, {Santana-Ros}, {Sarasso}, {Savietto}, {Schultheis}, {Sciacca},
		{Segol}, {Segovia}, {S{\'e}gransan}, {Shih}, {Siltala}, {Silva}, {Smart},
		{Smith}, {Solano}, {Solitro}, {Sordo}, {Soria Nieto}, {Souchay}, {Spagna},
		{Spoto}, {Stampa}, {Steele}, {Steidelm{\"u}ller}, {Stephenson}, {Stoev},
		{Suess}, {Surdej}, {Szabados}, {Szegedi-Elek}, {Tapiador}, {Taris}, {Tauran},
		{Taylor}, {Teixeira}, {Terrett}, {Teyssandier}, {Thuillot}, {Titarenko},
		{Torra Clotet}, {Turon}, {Ulla}, {Utrilla}, {Uzzi}, {Vaillant}, {Valentini},
		{Valette}, {van Elteren}, {Van Hemelryck}, {Vaschetto}, {Vecchiato},
		{Veljanoski}, {Viala}, {Vicente}, {Vogt}, {von Essen}, {Voss}, {Votruba},
		{Voutsinas}, {Walmsley}, {Weiler}, {Wertz}, {Wevers}, {Wyrzykowski},
		{Yoldas}, {{\v{Z}}erjal}, {Ziaeepour}, {Zorec}, {Zschocke}, {Zucker},
		{Zurbach}, \& {Zwitter}}]{2018A&A...616A..10G}
	{Gaia Collaboration}, {Babusiaux}, C., {van Leeuwen}, F., {et~al.} 2018, \aap,
	616, A10
	
	\bibitem[{{Gaia Collaboration} {et~al.}(2023){Gaia Collaboration}, {Vallenari},
		{Brown}, {Prusti}, {de Bruijne}, {Arenou}, {Babusiaux}, {Biermann},
		{Creevey}, {Ducourant}, {Evans}, {Eyer}, {Guerra}, {Hutton}, {Jordi},
		{Klioner}, {Lammers}, {Lindegren}, {Luri}, {Mignard}, {Panem}, {Pourbaix},
		{Randich}, {Sartoretti}, {Soubiran}, {Tanga}, {Walton}, {Bailer-Jones},
		{Bastian}, {Drimmel}, {Jansen}, {Katz}, {Lattanzi}, {van Leeuwen}, {Bakker},
		{Cacciari}, {Casta{\~n}eda}, {De Angeli}, {Fabricius}, {Fouesneau},
		{Fr{\'e}mat}, {Galluccio}, {Guerrier}, {Heiter}, {Masana}, {Messineo},
		{Mowlavi}, {Nicolas}, {Nienartowicz}, {Pailler}, {Panuzzo}, {Riclet}, {Roux},
		{Seabroke}, {Sordo}, {Th{\'e}venin}, {Gracia-Abril}, {Portell}, {Teyssier},
		{Altmann}, {Andrae}, {Audard}, {Bellas-Velidis}, {Benson}, {Berthier},
		{Blomme}, {Burgess}, {Busonero}, {Busso}, {C{\'a}novas}, {Carry}, {Cellino},
		{Cheek}, {Clementini}, {Damerdji}, {Davidson}, {de Teodoro}, {Nu{\~n}ez
			Campos}, {Delchambre}, {Dell'Oro}, {Esquej}, {Fern{\'a}ndez-Hern{\'a}ndez},
		{Fraile}, {Garabato}, {Garc{\'\i}a-Lario}, {Gosset}, {Haigron}, {Halbwachs},
		{Hambly}, {Harrison}, {Hern{\'a}ndez}, {Hestroffer}, {Hodgkin}, {Holl},
		{Jan{\ss}en}, {Jevardat de Fombelle}, {Jordan}, {Krone-Martins}, {Lanzafame},
		{L{\"o}ffler}, {Marchal}, {Marrese}, {Moitinho}, {Muinonen}, {Osborne},
		{Pancino}, {Pauwels}, {Recio-Blanco}, {Reyl{\'e}}, {Riello}, {Rimoldini},
		{Roegiers}, {Rybizki}, {Sarro}, {Siopis}, {Smith}, {Sozzetti}, {Utrilla},
		{van Leeuwen}, {Abbas}, {{\'A}brah{\'a}m}, {Abreu Aramburu}, {Aerts},
		{Aguado}, {Ajaj}, {Aldea-Montero}, {Altavilla}, {{\'A}lvarez}, {Alves},
		{Anders}, {Anderson}, {Anglada Varela}, {Antoja}, {Baines}, {Baker},
		{Balaguer-N{\'u}{\~n}ez}, {Balbinot}, {Balog}, {Barache}, {Barbato},
		{Barros}, {Barstow}, {Bartolom{\'e}}, {Bassilana}, {Bauchet}, {Becciani},
		{Bellazzini}, {Berihuete}, {Bernet}, {Bertone}, {Bianchi}, {Binnenfeld},
		{Blanco-Cuaresma}, {Blazere}, {Boch}, {Bombrun}, {Bossini}, {Bouquillon},
		{Bragaglia}, {Bramante}, {Breedt}, {Bressan}, {Brouillet}, {Brugaletta},
		{Bucciarelli}, {Burlacu}, {Butkevich}, {Buzzi}, {Caffau}, {Cancelliere},
		{Cantat-Gaudin}, {Carballo}, {Carlucci}, {Carnerero}, {Carrasco},
		{Casamiquela}, {Castellani}, {Castro-Ginard}, {Chaoul}, {Charlot}, {Chemin},
		{Chiaramida}, {Chiavassa}, {Chornay}, {Comoretto}, {Contursi}, {Cooper},
		{Cornez}, {Cowell}, {Crifo}, {Cropper}, {Crosta}, {Crowley}, {Dafonte},
		{Dapergolas}, {David}, {David}, {de Laverny}, {De Luise}, {De March}, {De
			Ridder}, {de Souza}, {de Torres}, {del Peloso}, {del Pozo}, {Delbo},
		{Delgado}, {Delisle}, {Demouchy}, {Dharmawardena}, {Di Matteo}, {Diakite},
		{Diener}, {Distefano}, {Dolding}, {Edvardsson}, {Enke}, {Fabre}, {Fabrizio},
		{Faigler}, {Fedorets}, {Fernique}, {Fienga}, {Figueras}, {Fournier},
		{Fouron}, {Fragkoudi}, {Gai}, {Garcia-Gutierrez}, {Garcia-Reinaldos},
		{Garc{\'\i}a-Torres}, {Garofalo}, {Gavel}, {Gavras}, {Gerlach}, {Geyer},
		{Giacobbe}, {Gilmore}, {Girona}, {Giuffrida}, {Gomel}, {Gomez},
		{Gonz{\'a}lez-N{\'u}{\~n}ez}, {Gonz{\'a}lez-Santamar{\'\i}a},
		{Gonz{\'a}lez-Vidal}, {Granvik}, {Guillout}, {Guiraud},
		{Guti{\'e}rrez-S{\'a}nchez}, {Guy}, {Hatzidimitriou}, {Hauser}, {Haywood},
		{Helmer}, {Helmi}, {Sarmiento}, {Hidalgo}, {Hilger}, {H{\l}adczuk}, {Hobbs},
		{Holland}, {Huckle}, {Jardine}, {Jasniewicz}, {Jean-Antoine Piccolo},
		{Jim{\'e}nez-Arranz}, {Jorissen}, {Juaristi Campillo}, {Julbe}, {Karbevska},
		{Kervella}, {Khanna}, {Kontizas}, {Kordopatis}, {Korn}, {K{\'o}sp{\'a}l},
		{Kostrzewa-Rutkowska}, {Kruszy{\'n}ska}, {Kun}, {Laizeau}, {Lambert},
		{Lanza}, {Lasne}, {Le Campion}, {Lebreton}, {Lebzelter}, {Leccia}, {Leclerc},
		{Lecoeur-Taibi}, {Liao}, {Licata}, {Lindstr{\o}m}, {Lister}, {Livanou},
		{Lobel}, {Lorca}, {Loup}, {Madrero Pardo}, {Magdaleno Romeo}, {Managau},
		{Mann}, {Manteiga}, {Marchant}, {Marconi}, {Marcos}, {Marcos Santos},
		{Mar{\'\i}n Pina}, {Marinoni}, {Marocco}, {Marshall}, {Martin Polo},
		{Mart{\'\i}n-Fleitas}, {Marton}, {Mary}, {Masip}, {Massari},
		{Mastrobuono-Battisti}, {Mazeh}, {McMillan}, {Messina}, {Michalik}, {Millar},
		{Mints}, {Molina}, {Molinaro}, {Moln{\'a}r}, {Monari}, {Mongui{\'o}},
		{Montegriffo}, {Montero}, {Mor}, {Mora}, {Morbidelli}, {Morel}, {Morris},
		{Muraveva}, {Murphy}, {Musella}, {Nagy}, {Noval}, {Oca{\~n}a}, {Ogden},
		{Ordenovic}, {Osinde}, {Pagani}, {Pagano}, {Palaversa}, {Palicio},
		{Pallas-Quintela}, {Panahi}, {Payne-Wardenaar}, {Pe{\~n}alosa Esteller},
		{Penttil{\"a}}, {Pichon}, {Piersimoni}, {Pineau}, {Plachy}, {Plum}, {Poggio},
		{Pr{\v{s}}a}, {Pulone}, {Racero}, {Ragaini}, {Rainer}, {Raiteri}, {Rambaux},
		{Ramos}, {Ramos-Lerate}, {Re Fiorentin}, {Regibo}, {Richards}, {Rios Diaz},
		{Ripepi}, {Riva}, {Rix}, {Rixon}, {Robichon}, {Robin}, {Robin}, {Roelens},
		{Rogues}, {Rohrbasser}, {Romero-G{\'o}mez}, {Rowell}, {Royer}, {Ruz Mieres},
		{Rybicki}, {Sadowski}, {S{\'a}ez N{\'u}{\~n}ez}, {Sagrist{\`a} Sell{\'e}s},
		{Sahlmann}, {Salguero}, {Samaras}, {Sanchez Gimenez}, {Sanna},
		{Santove{\~n}a}, {Sarasso}, {Schultheis}, {Sciacca}, {Segol}, {Segovia},
		{S{\'e}gransan}, {Semeux}, {Shahaf}, {Siddiqui}, {Siebert}, {Siltala},
		{Silvelo}, {Slezak}, {Slezak}, {Smart}, {Snaith}, {Solano}, {Solitro},
		{Souami}, {Souchay}, {Spagna}, {Spina}, {Spoto}, {Steele},
		{Steidelm{\"u}ller}, {Stephenson}, {S{\"u}veges}, {Surdej}, {Szabados},
		{Szegedi-Elek}, {Taris}, {Taylor}, {Teixeira}, {Tolomei}, {Tonello}, {Torra},
		{Torra}, {Torralba Elipe}, {Trabucchi}, {Tsounis}, {Turon}, {Ulla}, {Unger},
		{Vaillant}, {van Dillen}, {van Reeven}, {Vanel}, {Vecchiato}, {Viala},
		{Vicente}, {Voutsinas}, {Weiler}, {Wevers}, {Wyrzykowski}, {Yoldas}, {Yvard},
		{Zhao}, {Zorec}, {Zucker}, \& {Zwitter}}]{2023A&A...674A...1G}
	{Gaia Collaboration}, {Vallenari}, A., {Brown}, A.~G.~A., {et~al.} 2023, \aap,
	674, A1
	
	\bibitem[{{Ga{\l}an} {et~al.}(2012){Ga{\l}an}, {Miko{\l}ajewski}, {Tomov},
		{Graczyk}, {Apostolovska}, {Barzova}, {Bellas-Velidis}, {Bilkina}, {Blake},
		{Bolton}, {Bondar}, {Br{\'a}t}, {Bro{\.z}ek}, {Budzisz}, {Cika{\l}a},
		{Cs{\'a}k}, {Dapergolas}, {Dimitrov}, {Dobierski}, {Drahus},
		{Dr{\'o}{\.z}d{\.z}}, {Dvorak}, {Elder}, {Fr{\k{a}}ckowiak}, {Galazutdinov},
		{Gazeas}, {Georgiev}, {Gere}, {Go{\'z}dziewski}, {Grinin}, {Gromadzki},
		{Hajduk}, {Heras}, {Hopkins}, {Iliev}, {Janowski}, {Koci{\'a}n},
		{Ko{\l}aczkowski}, {Kolev}, {Kopacki}, {Krzesi{\'n}ski},
		{Ku{\v{c}}{\'a}kov{\'a}}, {Kuligowska}, {Kundera}, {Kurpi{\'n}ska-Winiarska},
		{Ku{\'z}micz}, {Liakos}, {Lister}, {Maciejewski}, {Majcher}, {Majewska},
		{Marrese}, {Michalska}, {Migaszewski}, {Miller}, {Munari}, {Musaev}, {Myers},
		{Narwid}, {N{\'e}meth}, {Niarchos}, {Niemczura}, {Og{\l}oza},
		{{\"O}{\v{g}}men}, {Oksanen}, {Osiwa{\l}a}, {Peneva}, {Pigulski}, {Popov},
		{Pych}, {Pye}, {Ragan}, {Roukema}, {R{\'o}{\.z}a{\'n}ski}, {Semkov}, {Siwak},
		{Staels}, {Stateva}, {Stempels}, {St{\c{e}}{\'s}licki},
		{{\'S}wierczy{\'n}ski}, {Szyma{\'n}ski}, {Tomov}, {Waniak}, {Wi{\c{e}}cek},
		{Winiarski}, {Wychudzki}, {Zajczyk}, {Zo{\l}a}, \&
		{Zwitter}}]{2012A&A...544A..53G}
	{Ga{\l}an}, C., {Miko{\l}ajewski}, M., {Tomov}, T., {et~al.} 2012, \aap, 544, A53
	
	\bibitem[{{Gibson} \& {Stencel}(2018)}]{2018MNRAS.476.5026G}
	{Gibson}, J.~L. \& {Stencel}, R.~E. 2018, \mnras, 476, 5026
	
	\bibitem[{{Green} {et~al.}(2019){Green}, {Schlafly}, {Zucker}, {Speagle}, \&
		{Finkbeiner}}]{2019ApJ...887...93G}
	{Green}, G.~M., {Schlafly}, E., {Zucker}, C., {Speagle}, J.~S., \&
	{Finkbeiner}, D. 2019, \apj, 887, 93
	
	\bibitem[{{Grinin} {et~al.}(2008){Grinin}, {Stempels}, {Gahm}, {Sergeev},
		{Arkharov}, {Barsunova}, \& {Tambovtseva}}]{2008A&A...489.1233G}
	{Grinin}, V., {Stempels}, H.~C., {Gahm}, G.~F., {et~al.} 2008, \aap, 489, 1233
	
	\bibitem[{{Heinze} {et~al.}(2018){Heinze}, {Tonry}, {Denneau}, {Flewelling},
		{Stalder}, {Rest}, {Smith}, {Smartt}, \& {Weiland}}]{2018AJ....156..241H}
	{Heinze}, A.~N., {Tonry}, J.~L., {Denneau}, L., {et~al.} 2018, \aj, 156, 241
	
	\bibitem[{{Hekker} {et~al.}(2011){Hekker}, {Gilliland}, {Elsworth}, {Chaplin},
		{De Ridder}, {Stello}, {Kallinger}, {Ibrahim}, {Klaus}, \&
		{Li}}]{2011MNRAS.414.2594H}
	{Hekker}, S., {Gilliland}, R.~L., {Elsworth}, Y., {et~al.} 2011, \mnras, 414,
	2594
	
	\bibitem[{{Kearns} \& {Herbst}(1998)}]{1998AJ....116..261K}
	{Kearns}, K.~E. \& {Herbst}, W. 1998, \aj, 116, 261
	
	\bibitem[{{Lallement} {et~al.}(2018){Lallement}, {Capitanio}, {Ruiz-Dern},
		{Danielski}, {Babusiaux}, {Vergely}, {Elyajouri}, {Arenou}, \&
		{Leclerc}}]{2018A&A...616A.132L}
	{Lallement}, R., {Capitanio}, L., {Ruiz-Dern}, L., {et~al.} 2018, \aap, 616,
	A132
	
	\bibitem[{{Mamajek} {et~al.}(2012){Mamajek}, {Quillen}, {Pecaut}, {Moolekamp},
		{Scott}, {Kenworthy}, {Collier Cameron}, \& {Parley}}]{2012AJ....143...72M}
	{Mamajek}, E.~E., {Quillen}, A.~C., {Pecaut}, M.~J., {et~al.} 2012, \aj, 143,
	72
	
	\bibitem[{{Masci} {et~al.}(2019){Masci}, {Laher}, {Rusholme}, {Shupe}, {Groom},
		{Surace}, {Jackson}, {Monkewitz}, {Beck}, {Flynn}, {Terek}, {Landry},
		{Hacopians}, {Desai}, {Howell}, {Brooke}, {Imel}, {Wachter}, {Ye}, {Lin},
		{Cenko}, {Cunningham}, {Rebbapragada}, {Bue}, {Miller}, {Mahabal}, {Bellm},
		{Patterson}, {Juri{\'c}}, {Golkhou}, {Ofek}, {Walters}, {Graham}, {Kasliwal},
		{Dekany}, {Kupfer}, {Burdge}, {Cannella}, {Barlow}, {Van Sistine}, {Giomi},
		{Fremling}, {Blagorodnova}, {Levitan}, {Riddle}, {Smith}, {Helou}, {Prince},
		\& {Kulkarni}}]{2019PASP..131a8003M}
	{Masci}, F.~J., {Laher}, R.~R., {Rusholme}, B., {et~al.} 2019, \pasp, 131,
	018003
	
	\bibitem[{{Mikul{\'a}{\v{s}}ek}(2015)}]{2015A&A...584A...8M}
	{Mikul{\'a}{\v{s}}ek}, Z. 2015, \aap, 584, A8
	
	\bibitem[{{Pickles}(1998)}]{1998PASP..110..863P}
	{Pickles}, A.~J. 1998, \pasp, 110, 863
	
	\bibitem[{{Pie{\'n}kowski} {et~al.}(2020){Pie{\'n}kowski}, {Ga{\l}an}, {Tomov},
		{Gazeas}, {Wychudzki}, {Miko{\l}ajewski}, {Kubicki}, {Staels}, {Zo{\l}a},
		{Pako{\'n}ska}, {D{\c{e}}bski}, {Kundera}, {Og{\l}oza}, {Dr{\'o}{\.z}d{\.z}},
		{Baran}, {Winiarski}, {Siwak}, {Dimitrov}, {Kjurkchieva}, {Marchev},
		{Armi{\'n}ski}, {Miller}, {Ko{\l}aczkowski}, {Mo{\'z}dzierski},
		{Zahajkiewicz}, {Bru{\'s}}, {Pigulski}, {Smela}, {Conseil}, {Boyd},
		{Conidis}, {Plauchu-Frayn}, {Heras}, {Kardasis}, {Biskupski}, {Kneip},
		{Hamb{\'a}lek}, {Pribulla}, {Kundra}, {Garai}, {Rodriguez}, {Kami{\'n}ski},
		{Dubois}, {Logie}, {Capetillo Blanco}, {Kankiewicz}, {{\'S}wierczy{\'n}ski},
		{Martignoni}, {Sergey}, {Kare Trandem Qvam}, {Semkov}, {Ibryamov}, {Peneva},
		{Gonzalez Carballo}, {Ribeiro}, {Dean}, {Apostolovska}, {Donchev}, {Corp},
		{McDonald}, {Rodriguez}, {Sanchez}, {Wiersema}, {Conseil}, {Menke}, {Sergey},
		\& {Richardson}}]{2020A&A...639A..23P}
	{Pie{\'n}kowski}, D., {Ga{\l}an}, C., {Tomov}, T., {et~al.} 2020, \aap, 639,
	A23
	
	\bibitem[{{Plavchan} {et~al.}(2008){Plavchan}, {Gee}, {Stapelfeldt}, \&
		{Becker}}]{2008ApJ...684L..37P}
	{Plavchan}, P., {Gee}, A.~H., {Stapelfeldt}, K., \& {Becker}, A. 2008, \apjl,
	684, L37
	
	\bibitem[{{Plavchan} {et~al.}(2013){Plavchan}, {G{\"u}th}, {Laohakunakorn}, \&
		{Parks}}]{2013A&A...554A.110P}
	{Plavchan}, P., {G{\"u}th}, T., {Laohakunakorn}, N., \& {Parks}, J.~R. 2013,
	\aap, 554, A110
	
	\bibitem[{{Poon} {et~al.}(2021){Poon}, {Zanazzi}, \&
		{Zhu}}]{2021MNRAS.503.1599P}
	{Poon}, M., {Zanazzi}, J.~J., \& {Zhu}, W. 2021, \mnras, 503, 1599
	
	\bibitem[{{Rodr{\'\i}guez-Ledesma} {et~al.}(2012){Rodr{\'\i}guez-Ledesma},
		{Mundt}, {Ibrahimov}, {Messina}, {Parihar}, {Hessman}, {Alves de Oliveira},
		\& {Herbst}}]{2012A&A...544A.112R}
	{Rodr{\'\i}guez-Ledesma}, M.~V., {Mundt}, R., {Ibrahimov}, M., {et~al.} 2012,
	\aap, 544, A112
	
	\bibitem[{{Rodr{\'\i}guez-Ledesma} {et~al.}(2013){Rodr{\'\i}guez-Ledesma},
		{Mundt}, {Pintado}, {Boudreault}, {Hessman}, \&
		{Herbst}}]{2013A&A...551A..44R}
	{Rodr{\'\i}guez-Ledesma}, M.~V., {Mundt}, R., {Pintado}, O., {et~al.} 2013,
	\aap, 551, A44
	
	\bibitem[{{Scott} {et~al.}(2014){Scott}, {Mamajek}, {Pecaut}, {Quillen},
		{Moolekamp}, \& {Bell}}]{2014ApJ...797....6S}
	{Scott}, E.~L., {Mamajek}, E.~E., {Pecaut}, M.~J., {et~al.} 2014, \apj, 797, 6
	
	\bibitem[{{Shingles} {et~al.}(2021){Shingles}, {Smith}, {Young}, {Smartt},
		{Tonry}, {Denneau}, {Heinze}, {Weiland}, {Flewelling}, {Stalder},
		{Clocchiatti}, {F{\"o}rster}, {Pignata}, {Rest}, {Anderson}, {Stubbs}, \&
		{Erasmus}}]{2021TNSAN...7....1S}
	{Shingles}, L., {Smith}, K.~W., {Young}, D.~R., {et~al.} 2021, Transient Name
	Server AstroNote, 7, 1
	
	\bibitem[{{Skrutskie} {et~al.}(2006){Skrutskie}, {Cutri}, {Stiening},
		{Weinberg}, {Schneider}, {Carpenter}, {Beichman}, {Capps}, {Chester},
		{Elias}, {Huchra}, {Liebert}, {Lonsdale}, {Monet}, {Price}, {Seitzer},
		{Jarrett}, {Kirkpatrick}, {Gizis}, {Howard}, {Evans}, {Fowler}, {Fullmer},
		{Hurt}, {Light}, {Kopan}, {Marsh}, {McCallon}, {Tam}, {Van Dyk}, \&
		{Wheelock}}]{2006AJ....131.1163S}
	{Skrutskie}, M.~F., {Cutri}, R.~M., {Stiening}, R., {et~al.} 2006, \aj, 131,
	1163
	
	\bibitem[{{Tonry} {et~al.}(2018){Tonry}, {Denneau}, {Heinze}, {Stalder},
		{Smith}, {Smartt}, {Stubbs}, {Weiland}, \& {Rest}}]{2018PASP..130f4505T}
	{Tonry}, J.~L., {Denneau}, L., {Heinze}, A.~N., {et~al.} 2018, \pasp, 130,
	064505
	
	\bibitem[{{van Dam} {et~al.}(2020){van Dam}, {Kenworthy}, {David}, {Mamajek},
		{Hillenbrand}, {Cody}, {Howard}, {Isaacson}, {Ciardi}, {Rebull}, {Stauffer},
		{Patel}, {Cameron + WASP Collaborators}, {Rodriguez}, {Pojma{\'n}ski},
		{Gonzales}, {Schlieder}, {Hambsch}, {Dufoer}, {Vanmunster}, {Dubois},
		{Vanaverbeke}, {Logie}, \& {Rau}}]{2020AJ....160..285V}
	{van Dam}, D.~M., {Kenworthy}, M.~A., {David}, T.~J., {et~al.} 2020, \aj, 160,
	285
	
	\bibitem[{{Wang} \& {Chen}(2019)}]{2019ApJ...877..116W}
	{Wang}, S. \& {Chen}, X. 2019, \apj, 877, 116
	
	\bibitem[{{Wright} {et~al.}(2010){Wright}, {Eisenhardt}, {Mainzer}, {Ressler},
		{Cutri}, {Jarrett}, {Kirkpatrick}, {Padgett}, {McMillan}, {Skrutskie},
		{Stanford}, {Cohen}, {Walker}, {Mather}, {Leisawitz}, {Gautier}, {McLean},
		{Benford}, {Lonsdale}, {Blain}, {Mendez}, {Irace}, {Duval}, {Liu}, {Royer},
		{Heinrichsen}, {Howard}, {Shannon}, {Kendall}, {Walsh}, {Larsen}, {Cardon},
		{Schick}, {Schwalm}, {Abid}, {Fabinsky}, {Naes}, \&
		{Tsai}}]{2010AJ....140.1868W}
	{Wright}, E.~L., {Eisenhardt}, P. R.~M., {Mainzer}, A.~K., {et~al.} 2010, \aj,
	140, 1868
	
	\bibitem[{{Zhu} {et~al.}(2022){Zhu}, {Bernhard}, {Dai}, {Fang}, {Zanazzi},
		{Zang}, {Dong}, {Hambsch}, {Gan}, {Wu}, \& {Poon}}]{2022ApJ...933L..21Z}
	{Zhu}, W., {Bernhard}, K., {Dai}, F., {et~al.} 2022, \apjl, 933, L21
	
\end{thebibliography}

\listofobjects
\end{document}